\documentstyle[12pt]{article}
\newlength{\extraspace}
\newlength{\extraspaces}
\newcommand{\be}{\begin{equation}
\addtolength{\abovedisplayskip}{\extraspaces}
\addtolength{\belowdisplayskip}{\extraspaces}
\addtolength{\abovedisplayshortskip}{\extraspace}
\addtolength{\belowdisplayshortskip}{\extraspace}}
\newcommand{\ee}{\end{equation}}
\newcommand{\ba}{\begin{eqnarray}
\addtolength{\abovedisplayskip}{\extraspaces}
\addtolength{\belowdisplayskip}{\extraspaces}
\addtolength{\abovedisplayshortskip}{\extraspace}
\addtolength{\belowdisplayshortskip}{\extraspace}}
\newcommand{\ea}{\end{eqnarray}}
\newcommand{\nonu}{\nonumber \\[.5mm]}
\newcommand{\A}{&\!\!\!}
\begin{document}
\thispagestyle{empty}
\begin{flushright}
SIT-LP-02/06 \\
{\tt hep-th/0206122} \\
June, 2002
\end{flushright}
\vspace{7mm}
\begin{center}
{\large \bf Linearization of $N = 2$ Nonlinear Supersymmetry \\
and Spontaneous Supersymmetry Breaking Parameters 
} \\[20mm]
{\sc Kazunari Shima}
\footnote{
\tt e-mail: shima@sit.ac.jp} \ 
and \ 
{\sc Motomu Tsuda}
\footnote{
\tt e-mail: tsuda@sit.ac.jp} 
\\[5mm]
{\it Laboratory of Physics, 
Saitama Institute of Technology \\
Okabe-machi, Saitama 369-0293, Japan} \\[20mm]
\begin{abstract}
We show explicitly by the heuristic and practical arguments 
that for $N = 2$ supersymmetry (SUSY) 
a SUSY invariant relation between component fields 
of a vector supermultiplet of linear SUSY 
and Nambu-Goldstone fermions of the Volkov-Akulov model 
of nonlinear SUSY is written by using only three arbitrary 
dimensionless parameters, which can be recasted 
as the vacuum expectation values of auxiliary fields 
in the vector supermultiplet. 
\end{abstract}
\end{center}

\newpage

Spontaneously breaking of supersymmetry (SUSY) \cite{WZ1}-\cite{VA} 
gives rise to Nambu-Goldstone (N-G) fermions \cite{SS}-\cite{O} 
which can be characterized by means of the nonlinear 
realization of global SUSY (NL SUSY) \cite{VA}. 
The relationship between the Volkov-Akulov (V-A) model 
of NL SUSY and a scalar supermultiplet of the linear SUSY 
of Wess and Zumino \cite{WZ1} is well-known 
from the early work by many authors \cite{IK}-\cite{Nishi}. 
As for a vector supermultiplet, its relationship to $N = 1$ V-A model 
was studied in the context of the coupling of the V-A action 
to the gauge multiplet action with the Fayet-Iliopoulos $D$ term 
of linear SUSY \cite{IK}. 

Recently, it becomes important to linearize NL SUSY 
from the viewpoint of the superon-graviton 
model (SGM) having a new NL SUSY based upon SO(10) super-Poincar\'e 
algebra from a composite viewpoint of matter proposed 
by one of the author \cite{KS1,KS2}. 
The fundamental action of SGM is the Einstein-Hilbert type one 
which is invariant under at least 
$[{\rm global\ NL\ SUSY}] \otimes [{\rm local\ GL(4,R)}] 
\otimes [{\rm local\ Lorentz}] \otimes [{\rm global\ SO(N)}]$ 
as a whole \cite{ST1}, which is isomorphic to 
global SO(N) super-Poincar\'e symmetry. 
The expansion of the SGM action in terms of graviton and superons 
(N-G fermions) with spin-1/2 is a highly nonlinear one 
which consists of the Einstein-Hilbert action of general relativity, 
the V-A action and their interactions \cite{ST2}. 
In SGM, the (composite) eigenstates of the {\it linear} representation 
of SO(10) super-Poincar\'e algebra which is composed of superons 
(N-G fermions) with spin-1/2 are regarded as all observed elementary 
particles at low energy except graviton \cite{KS1,KS2}. 
Therefore it is inevitable to linearize such a highly nonlinear theory 
for deriving the low energy physical contents of the SGM action, 
and so the investigation of the linearization of V-A model in detail 
is also useful. 

As a preliminary to do this, in Ref.\cite{STT1} we have explicitly shown 
that the $N = 1$ V-A model is related to the total action 
of a U(1) gauge supermultiplet \cite{WZ2} 
of the linear SUSY with the Fayet-Iliopoulos $D$ term indicating 
a spontaneous SUSY breaking. 
In this work of \cite{STT1} it became clear that the representations of 
all component fields of a U(1) gauge supermultiplet 
in terms of the N-G fermion indicate 
the axial vector nature of the U(1) gauge field. 
Furthermore, in \cite{STT2} we have constructed for $N = 2$ SUSY 
a relation between component fields 
of a vector supermultiplet \cite{Fa} and N-G fermion fields 
at the leading orders in a SUSY invariant way, 
and we have shown that the U(1) gauge field has a (realistic) vector 
nature in terms of the N-G fermion fields. 
In this paper, starting from an general ansatz 
given below (Eq.(\ref{ansatz})), 
we show explicitly that the SUSY invariant relation 
of the $N = 2$ theories is written by using only three arbitrary 
dimensionless parameters, which can be recasted 
as the vacuum expectation values of auxiliary fields 
in a vector supermultiplet. 

We denote in this paper the component fields of an $N = 2$ U(1) 
gauge supermultiplet \cite{Fa} as follows; 
\footnote{
In this letter Minkowski spacetime indices are denoted 
by $a, b, ... = 0, 1, 2, 3$, 
and we use the  Minkowski spacetime metric 
${1 \over 2}\{ \gamma^a, \gamma^b \} = \eta^{ab}= (+, -, -, -)$ 
and $\sigma^{ab} = {i \over 4}[\gamma^a, \gamma^b]$.}
namely, $A$ and $B$ for two physical scalar fields, 
$A_a$ for a U(1) gauge field 
and $\lambda^i \ (i = 1, 2)$ for two Majorana spinors 
in addition to $F$, $G$ and $D$ for three auxiliary scalar fields 
at least for a free vector supermultiplet 
required from the mismatch of the degrees of freedom 
between bosonic and fermionic physical fields. 
\footnote{The component fields indeed belong to representations 
of a rigid SU(2) \cite{Fa}; namely, $\lambda^i$ and ($F$, $G$, $D$) 
belong to representations {\bf 2} and {\bf 3} of SU(2) 
respectively while other fields are singlets. 
In Ref.\cite{STT2} we linearize $N = 2$ NL SUSY 
in the manifestly SU(2) covariant form.} 
The linear SUSY transformations of these component fields 
generated by constant (Majorana) spinor parametars $\zeta^i$ are 
\ba
\A \A 
\delta A = \bar\zeta^1 \lambda^1 + \bar\zeta^2 \lambda^2, \nonu
\A \A 
\delta B = i \bar\zeta^1 \gamma_5 \lambda^1 
+ i \bar\zeta^2 \gamma_5 \lambda^2, \nonu
\A \A 
\delta A_a = - i \bar\zeta^1 \gamma_a \lambda^2 
+ i \bar\zeta^2 \gamma_a \lambda^1, \nonu
\A \A 
\delta \lambda^1 = \{ (F + i \gamma_5 G) 
- i \!\!\not\!\partial (A + i \gamma_5 B) \} \zeta^1 
- i F_{ab} \sigma^{ab} \zeta^2 + i \gamma_5 \zeta^2 D, \nonu
\A \A 
\delta \lambda^2 = \{ (F - i \gamma_5 G) 
- i \!\!\not\!\partial (A + i \gamma_5 B) \} \zeta^2 
+ i F_{ab} \sigma^{ab} \zeta^1 + i \gamma_5 \zeta^1 D, \nonu
\A \A 
\delta F = - i \bar\zeta^1 \!\!\not\!\partial \lambda^1 
- i \bar\zeta^2 \!\!\not\!\partial \lambda^2, \nonu
\A \A 
\delta G = \bar\zeta^1 \gamma_5 \!\!\not\!\partial \lambda^1 
- \bar\zeta^2 \gamma_5 \!\!\not\!\partial \lambda^2, \nonu
\A \A 
\delta D = \bar\zeta^1 \gamma_5 \!\!\not\!\partial \lambda^2 
+ \bar\zeta^2 \gamma_5 \!\!\not\!\partial \lambda^1, 
\label{LSUSY}
\ea
which satisfy a closed off-shell commutator algebra. 

On the other hand, in the $N = 2$ V-A model \cite{BV} 
we have a NL SUSY transformation laws of (Majorana) N-G fermions 
$\psi^i$ generated by $\zeta^i$, 
\be
\delta \psi^i = {1 \over \kappa} \zeta^i 
- i \kappa (\bar\zeta^j \gamma^a \psi^j) \partial_a \psi^i, 
\label{NLSUSY}
\ee
where $\kappa$ is a constant whose dimension is $({\rm mass})^{-2}$. 
Eq.(\ref{NLSUSY}) also satisfies the off-shell commutator algebra without 
a U(1) gauge transformation. 

From the SUSY transformations (\ref{LSUSY}) and (\ref{NLSUSY}), 
a SUSY invariant relation between the component fields 
of the $N = 2$ vector supermultiplet and the N-G fermion fields 
$\psi^i$ is obtained at the leading orders of $\kappa$ 
as follows: Indeed, adopting an ansatz 
\ba
\A \A \lambda^1 = (\xi + i \theta \gamma_5) \psi^1 
+ (\eta + i \varphi \gamma_5) \psi^2 + ...\ , \nonu
\A \A \lambda^2 = (\xi' + i \theta' \gamma_5) \psi^1 
+ (\eta' + i \varphi' \gamma_5) \psi^2 + ...\ . 
\label{ansatz}
\ea
with $\xi, \eta, \theta, \varphi, \xi', \eta', \theta'$ 
and $\varphi'$ being eight arbitrary real parameters 
which are the most general one for the dimensionless case, 
we substitute (\ref{ansatz}) into (\ref{LSUSY}) 
and combine (\ref{LSUSY}) with (\ref{NLSUSY}) as done in \cite{R}. 
Then we immediately obtain the relation between the bosonic fields 
$A$, $B$, $A_a$, $F$, $G$ and $D$ of the linear supermultiplet 
and the N-G fermions $\psi^i$ at the leading orders of $\kappa$, 
provided that the real parameters in (\ref{ansatz}) 
are restricted as 
\ba
\A \A \eta = 0 = \xi', \nonu
\A \A \eta' = \xi, \ \ \ \theta' = \varphi, 
      \ \ \ \varphi' = - \theta. 
\ea
The results are 
\ba
\A \A 
A = {1 \over 2} \kappa \ \xi \ (\bar\psi^1 \psi^1 + \bar\psi^2 \psi^2) 
+ {i \over 2} \kappa \ \theta 
\ (\bar\psi^1 \gamma_5 \psi^1 - \bar\psi^2 \gamma_5 \psi^2) 
+ i \kappa \ \varphi \ \bar\psi^1 \gamma_5 \psi^2 + ...\ , 
\label{co-A} \\
\A \A 
B = {i \over 2} \kappa \ \xi \ (\bar\psi^1 \gamma_5 \psi^1 
+ \bar\psi^2 \gamma_5 \psi^2) 
- {1 \over 2} \kappa \ \theta \ (\bar\psi^1 \psi^1 - \bar\psi^2 \psi^2) 
- \kappa \ \varphi \ \bar\psi^1 \psi^2 + ...\ , \\
\A \A 
A_a = - i \kappa \ \xi \ \bar\psi^1 \gamma_a \psi^2 
+ \kappa \ \theta \ \bar\psi^1 \gamma_5 \gamma_a \psi^2 
- {1 \over 2} \kappa \ \varphi \ (\bar\psi^1 \gamma_5 \gamma_a \psi^1 
- \bar\psi^2 \gamma_5 \gamma_a \psi^2) + ...\ , 
\label{co-Aa} \\
\A \A 
\lambda^1 = (\xi + i \theta \gamma_5) \psi^1 
+ i \varphi \gamma_5 \psi^2 + ...\ , 
\label{co-l11} \\
\A \A 
\lambda^2 = (\xi - i \theta \gamma_5) \psi^2 
+ i \varphi \gamma_5 \psi^1 + ...\ , 
\label{co-l21} \\
\A \A 
F = \xi \ \left\{ {1 \over \kappa} 
- i \kappa (\bar\psi^1 \!\!\not\!\partial \psi^1 
+ \bar\psi^2 \!\!\not\!\partial \psi^2) \right\} 
- \kappa \ \theta \ (\bar\psi^1 \gamma_5 \!\!\not\!\partial \psi^1 
- \bar\psi^2 \gamma_5 \!\!\not\!\partial \psi^2) \nonu
\A \A \hspace{8mm} 
- \kappa \ \varphi \ \partial_a (\bar\psi^1 \gamma_5 \gamma^a \psi^2) 
+ ...\ , 
\label{co-F} \\
\A \A 
G = \theta \ \left\{ {1 \over \kappa} 
- i \kappa (\bar\psi^1 \!\!\not\!\partial \psi^1 
+ \bar\psi^2 \!\!\not\!\partial \psi^2) \right\} 
+ \kappa \ \xi \ (\bar\psi^1 \gamma_5 \!\!\not\!\partial \psi^1 
- \bar\psi^2 \gamma_5 \!\!\not\!\partial \psi^2) \nonu
\A \A \hspace{8mm} 
- i \kappa \ \varphi \ \partial_a (\bar\psi^1 \gamma^a \psi^2) + ...\ , \\
\A \A 
D = \varphi \ \left\{ {1 \over \kappa} 
- i \kappa (\bar\psi^1 \!\!\not\!\partial \psi^1 
+ \bar\psi^2 \!\!\not\!\partial \psi^2) \right\} 
+ \kappa \ \xi \ \partial_a (\bar\psi^1 \gamma_5 \gamma^a \psi^2) \nonu
\A \A \hspace{8mm} 
+ i \kappa \ \theta \ \partial_a (\bar\psi^1 \gamma^a \psi^2) + ...\ , 
\label{co-D}
\ea
in which the three arbitrary real parameters $\xi$, $\theta$ 
and $\varphi$ are involved. The first term 
$- i \kappa \ \xi \ \bar\psi^1 \gamma_a \psi^2$ 
in Eq.(\ref{co-Aa}) shows 
the vector nature of the U(1) gauge field 
as we expected and have already showed \cite{STT2}. 
Also Eqs.(\ref{co-F}) to (\ref{co-D}) 
for the auxiliary fields $F$, $G$ and $D$ 
have the form which is proportional to a determinant 
$\vert w \vert = {\rm det}(w{^a}_b)$ in the $N = 2$ V-A model 
\cite{BV} with $w{^a}_b$ being defined by 
\be
w{^a}_b = \delta{^a}_b + t{^a}_b, \ \ \ 
t{^a}_b = - i \kappa^2 \bar\psi^i \gamma^a \partial_b \psi^i, 
\ee
plus total derivative terms at least at the leading 
orders of $\kappa$. Namely, we have 
\ba
\A \A 
F = {\xi \over \kappa} \ [ \ {\rm leading\ terms\ of\ } \vert w \vert \ ] 
+ [ \ {\rm tot.\ der.} \ ] + ...\ , \\
\A \A 
G = {\theta \over \kappa} \ [ \ {\rm leading\ terms\ of\ } \vert w \vert \ ] 
+ [ \ {\rm tot.\ der.} \ ] + ...\ , \\
\A \A 
D = {\varphi \over \kappa} \ [ \ {\rm leading\ terms\ of\ } \vert w \vert \ ] 
+ [ \ {\rm tot.\ der.} \ ] + ...\ . 
\ea
In addition, the first terms in Eqs.(\ref{co-F}) to (\ref{co-D}) 
or the SUSY transformations of Eqs.(\ref{co-l11}) and (\ref{co-l21}) 
show that $\xi/\kappa$, $\theta/\kappa$ and $\varphi/\kappa$ 
correspond to the vacuum expectation values 
of the auxiliary fields $F$, $G$ and $D$. 

We can continue to obtain higher order terms in the SUSY invarinat 
relations: After some calculations we obtain the relation between 
$\lambda^i$ and the N-G fermion fields $\psi^i$ at $O(\kappa^2)$ as 
\ba
\lambda^1 = \A \A 
(\xi + i \theta \gamma_5) \psi^1 + i \varphi \gamma_5 \psi^2 \nonu
\A \A 
- \ {i \over 2} \kappa^2 \ \xi 
\ \{ (\bar\psi^1 \!\!\not\!\partial \psi^1) \psi^1 
- (\bar\psi^1 \gamma_5 \!\!\not\!\partial \psi^1) \gamma_5 \psi^1 
+ (\bar\psi^1 \partial_a \psi^1) \gamma^a \psi^1 
+ (\bar\psi^1 \gamma_5 \partial_a \psi^1) \gamma_5 \gamma^a \psi^1 \} \nonu
\A \A 
- \ {1 \over 2} \kappa^2 \ \theta 
\ \{ (\bar\psi^1 \gamma_5 \!\!\not\!\partial \psi^1) \psi^1 
- (\bar\psi^1 \!\!\not\!\partial \psi^1) \gamma_5 \psi^1 
- (\bar\psi^1 \gamma_5 \partial_a \psi^1) \gamma^a \psi^1 
- (\bar\psi^1 \partial_a \psi^1) \gamma_5 \gamma^a \psi^1 \} \nonu
\A \A 
- \ {1 \over 2} \kappa^2 \ \varphi 
\ \{ (\bar\psi^1 \gamma_5 \!\!\not\!\partial \psi^2) \psi^1 
- (\bar\psi^1 \!\!\not\!\partial \psi^2) \gamma_5 \psi^1 
- (\bar\psi^1 \gamma_5 \partial_a \psi^2) \gamma^a \psi^1 
- (\bar\psi^1 \partial_a \psi^2) \gamma_5 \gamma^a \psi^1 \} \nonu
\A \A 
- \ i \kappa^2 \ \xi 
\ \{ (\bar\psi^2 \!\!\not\!\partial \psi^2) \psi^1 
+ (\bar\psi^2 \gamma_5 \!\!\not\!\partial \psi^2) \gamma_5 \psi^1 
+ (\bar\psi^2 \partial_a \psi^2) \gamma^a \psi^1 
+ (\bar\psi^2 \gamma_5 \partial_a \psi^2) \gamma_5 \gamma^a \psi^1 \} \nonu
\A \A 
+ \ \kappa^2 \ \theta 
\ \{ (\bar\psi^2 \gamma_5 \!\!\not\!\partial \psi^2) \psi^1 
+ (\bar\psi^2 \!\!\not\!\partial \psi^2) \gamma_5 \psi^1 
- (\bar\psi^2 \gamma_5 \partial_a \psi^2) \gamma^a \psi^1 
- (\bar\psi^2 \partial_a \psi^2) \gamma_5 \gamma^a \psi^1 \} \nonu
\A \A 
- \ \kappa^2 \ \varphi 
\ \{ (\bar\psi^2 \gamma_5 \!\!\not\!\partial \psi^1) \psi^1 
+ (\bar\psi^2 \!\!\not\!\partial \psi^1) \gamma_5 \psi^1 
- (\bar\psi^2 \gamma_5 \partial_a \psi^1) \gamma^a \psi^1 
- (\bar\psi^2 \partial_a \psi^1) \gamma_5 \gamma^a \psi^1 \} \nonu
\A \A 
+ \ \kappa^2 \ \xi \left\{ 
{i \over 2} (\bar\psi^2 \gamma_5 \!\!\not\!\partial \psi^1) \gamma_5 \psi^2 
+ (\bar\psi^2 \gamma_b \partial_a \psi^1) \sigma^{ab} \psi^2 \right\} \nonu
\A \A 
+ \ \kappa^2 \ \theta \left\{ 
{1 \over 2} (\bar\psi^2 \!\!\not\!\partial \psi^1) \gamma_5 \psi^2 
- i (\bar\psi^2 \gamma_5 \gamma_b \partial_a \psi^1) 
\sigma^{ab} \psi^2 \right\} \nonu
\A \A 
+ \ \kappa^2 \ \varphi \left\{ 
{1 \over 2} (\bar\psi^2 \!\!\not\!\partial \psi^2) \gamma_5 \psi^2 
- i (\bar\psi^2 \gamma_5 \gamma_b \partial_a \psi^2) 
\sigma^{ab} \psi^2 \right\} + ...\ , 
\label{co-l12}
\ea
and $\lambda^2$ is obtained by exchanging the indices 1 and 2 
and by replacing $\theta$ with $-\theta$ in Eq.(\ref{co-l12}). 
We can also construct the SUSY invariant relation 
with respect to the bosonic fields of the linear supermultiplet 
at $O(\kappa^3)$ \cite{STT2}. 
In principle we can further continue to obtain higher order terms 
in the SUSY invariant relation following this approach. 
However, it will be more useful to use the $N = 2$ superfield formalism 
\cite{GSW} as was done in Refs.\cite{IK,UZ,STT1}. 
Remarkably, Eqs.(\ref{co-A}) to (\ref{co-D}) 
(and also (\ref{co-l12}), etc.) 
reduce to that of the $N = 1$ SUSY by imposing, e.g. $\psi^2 = 0$: 
Indeed, when $\xi = 1$ and $\theta = \varphi = 0$, 
they becomes that of the scalar supermultiplet obtained in Ref.\cite{R}. 
When $\varphi = 1$ and $\xi = \theta = 0$, 
they reduce to that of the U(1) gauge supermultiplet obtained 
in Refs.\cite{IK,STT1}. 

So far, our discussion does not depend on the form 
of the action for the two models. 
We now consider a free action which is invariant under Eq.(\ref{LSUSY}) 
\ba
S_{\rm lin} = \A \A \int d^4 x \left[ {1 \over 2} (\partial_a A)^2 
+ {1 \over 2} (\partial_a B)^2 - {1 \over 4} F^2_{ab} 
+ {i \over 2} \bar\lambda^i \!\!\not\!\partial \lambda^i
+ {1 \over 2} (F^2 + G^2 + D^2) \right. \nonu
\A \A 
\hspace{1.2cm} \left. 
- {1 \over \kappa} (\xi F + \theta G + \varphi D) \right], 
\label{Lact}
\ea
where $\xi$, $\theta$ and $\varphi$ are three arbitraty 
real parameters satisfying $\xi^2 + \theta^2 + \varphi^2 = 1$. 
The last three terms proportional to $\kappa^{-1}$ 
is an analog of the Fayet-Iliopoulos D term in the $N = 1$ 
theories \cite{FI}. 
The field equations for the auxiliary fields $F$, $G$ or $D$ 
are $F = \xi/\kappa$, $G = \theta/\kappa$ 
or $D = \varphi/\kappa$ indicating a spontaneous SUSY breaking. 
Substituting (\ref{co-A}) to (\ref{co-D}) 
into the linear action $S_{\rm lin}$ of (\ref{Lact}), 
we can show immediately that $S_{\rm lin}$ coincides with 
the following V-A action $S_{\rm VA}$  up to and including $O(\kappa^0)$; 
namely, 
\ba
S_{\rm VA} = \A \A - {1 \over {2 \kappa^2}} 
\int d^4 x \ \vert w \vert \nonu
= \A \A 
- {1 \over {2 \kappa^2}} \int d^4 x 
\left[ 1 + t{^a}_a 
+ {1 \over 2}(t{^a}_a t{^b}_b - t{^a}_b t{^b}_a) \right. \nonu
\A \A 
\left. - {1 \over 6} \epsilon_{abcd} \epsilon^{efgd} t{^a}_e t{^b}_f t{^c}_g 
- {1 \over 4!} \epsilon_{abcd} \epsilon^{efgh} t{^a}_e t{^b}_f t{^c}_g t{^d}_h 
\right], 
\label{VAact}
\ea
which is invariant under (\ref{NLSUSY}). 

We note that the linearization of $N = 2$ SUSY in this paper 
can be discussed as a manifestly (rigid) SU(2) invariant form 
\cite{STT2}, which gives more concise expressions of the SUSY 
invariant relation (\ref{co-A}) to (\ref{co-D}) 
(and also (\ref{co-l12}), etc.): 
By defining 
\ba
\A \A \lambda_{Li} = {1 \over 2} (1 - \gamma_5) \lambda^i, \qquad 
      \lambda_R^i = {1 \over 2} (1 + \gamma_5) \lambda^i, \nonu
\A \A \zeta_L^i = {1 \over 2} (1 - \gamma_5) \zeta^i, \qquad 
      \zeta_{Ri} = {1 \over 2}(1 + \gamma_5) \zeta^i, 
\ea
and their Dirac conjugates $\bar\lambda_L^i$, $\bar\lambda_{Ri}$, 
$\bar\zeta_{Li}$ and $\bar\zeta_R^i$, 
and by using the antisymmetric symbol 
$\epsilon^{ij}$ ($\epsilon^{12} = +1$) and 
$\epsilon_{ij}$ ($\epsilon_{12} = -1$) to raise and lower SU(2) 
indices as $\psi^i = \epsilon^{ij} \psi_j$, 
$\psi_i = \epsilon_{ij} \psi^j$, 
the linear SUSY transformations of (\ref{LSUSY}) are 
rewritten as 
\ba
\delta A \A = \A \bar\zeta_{Li} \lambda_R^i 
+ \bar\zeta_R^i \lambda_{Li}, \nonu
\delta B \A = \A i \bar\zeta_{Li} \lambda_R^i 
- i \bar\zeta_R^i \lambda_{Li}, \nonu
\delta A_a \A = \A - i \bar\zeta_{Li} \gamma_a \lambda_L^i 
+ i \bar\zeta_R^i \lambda_{Ri}, \nonu
\delta \lambda_{Li} \A = \A i F_{ab} \sigma^{ab} \zeta_{Li} 
- i \!\!\not\!\partial (A + iB) \zeta_{Ri} 
+ i \zeta_{Lj} (\sigma^I)^j{}_i D^I, \nonu
\delta \lambda_R^i \A = \A - i F_{ab} \sigma^{ab} \zeta_R^i 
- i \!\!\not\!\partial (A - iB) \zeta_L^i 
+ i D^I (\sigma^I)^i{}_j \zeta_R^j, \nonu
\delta D^I \A = \A \bar\zeta_{Li} (\sigma^I)^i{}_j 
\!\!\not\!\partial \lambda_L^j 
+ \bar\zeta_{Ri} (\sigma^I)^i{}_j \!\!\not\!\partial \lambda_R^j, 
\ea
where $\sigma^I$ are the Pauli matrices and $D^I$ are SU(2) vector 
defined by 
\be
D^1 = - G, \qquad D^2= - F, \qquad D^3 = D. 
\ee
And, for example, Eq.(\ref{co-l12}) is recasted as the following 
compact form; namely, 
\ba
\lambda_{Li}(\psi) \A = \A i \xi^I (\psi_L \sigma^I)_i 
+ \kappa^2 \xi^I \gamma^a \psi_{Ri} \bar\psi_R \sigma^I 
\partial_a \psi_L 
+ {1 \over 2} \kappa^2 \xi^I \gamma^{ab} \psi_{Li} 
\partial_a \left( \bar\psi_L \sigma^I \gamma_b \psi_L \right) \nonu
\A\A + {1 \over 2} \kappa^2 \xi^I ( \psi_L \sigma^J )_i 
\left( \bar\psi_L \sigma^J \sigma^I \!\!\not\!\partial \psi_L 
- \bar\psi_R \sigma^J \sigma^I \!\!\not\!\partial \psi_R \right) 
+ {\cal O}(\kappa^4), 
\ea

In conclusion of this paper, adopting the general ansatz (\ref{ansatz}) 
having the eight real dimensionless parameters with $\kappa^0$, 
we have explicitly shown that for $N = 2$ SUSY the SUSY invariant relation 
(\ref{co-A}) to (\ref{co-D}) (and also (\ref{co-l12}), etc.) 
is written by using only three arbitrary parameters, 
which can be recasted as the vacuum expectation values 
of the auxiliary fields in the vector supermultiplet. 
These heuristic arguments are practical and show more general assumptions 
adopted for obtaining the SUSY invariant relation. 
From those arguments on the linearization of $N = 1$ and $N = 2$ SUSY, 
we speculate that all renormalizable ($N$-exteded) global linear SUSY 
(interacting) model are equivalent to the ($N$-extended) V-A model. 
It is remarkable that the vector (not axial) gauge field 
needs $N = 2$ SUSY, i.e. SU(2) structure of the basic algebra. 
These results may support the SGM scenario \cite{KS1,KS2} 
which is global NL SUSY (generalized V-A) model in curved spacetime.

\vspace{10mm}

\noindent {\Large{\bf Acknowledgements}}

\vspace{3mm}

The work of M.T. is supported in part by the High-Tech research 
program of Saitama Institute of Technology. 

\newpage

%
\newcommand{\NP}[1]{{\it Nucl.\ Phys.\ }{\bf #1}}
\newcommand{\PL}[1]{{\it Phys.\ Lett.\ }{\bf #1}}
\newcommand{\CMP}[1]{{\it Commun.\ Math.\ Phys.\ }{\bf #1}}
\newcommand{\MPL}[1]{{\it Mod.\ Phys.\ Lett.\ }{\bf #1}}
\newcommand{\IJMP}[1]{{\it Int.\ J. Mod.\ Phys.\ }{\bf #1}}
\newcommand{\PR}[1]{{\it Phys.\ Rev.\ }{\bf #1}}
\newcommand{\PRL}[1]{{\it Phys.\ Rev.\ Lett.\ }{\bf #1}}
\newcommand{\PTP}[1]{{\it Prog.\ Theor.\ Phys.\ }{\bf #1}}
\newcommand{\PTPS}[1]{{\it Prog.\ Theor.\ Phys.\ Suppl.\ }{\bf #1}}
\newcommand{\AP}[1]{{\it Ann.\ Phys.\ }{\bf #1}}

\end{document}